\documentstyle[12pt]{article}

\begin{document}

\baselineskip 6mm

\begin{titlepage}

%---------------- preprint number & date ---------------
\hfill\parbox{4cm}
{ KIAS-P98019 \\ hep-th/9808183 \\ August 1998}

%------------------------ title ------------------------
\vspace{15mm}
\begin{center}
{\Large \bf
Effective Action for Membrane Dynamics in DLCQ $M$ theory
on a Two-torus
}
\end{center}

%---------------- authors and addresses ----------------
\vspace{5mm}
\begin{center} 
Seungjoon Hyun\footnote{\tt hyun@kiasph.kaist.ac.kr},
Youngjai Kiem\footnote{\tt ykiem@kiasph.kaist.ac.kr},
and Hyeonjoon Shin\footnote{\tt hshin@kiasph.kaist.ac.kr}
\\[5mm]
{\it 
School of Physics, Korea Institute for Advanced Study,
Seoul 130-012, Korea
}
\end{center}
\thispagestyle{empty}

%----------------------- abstract ----------------------
\vfill
\begin{center}
{\bf Abstract}
\end{center}
\noindent
The effective action for the membrane dynamics on the background
geometry of the $N$-sector DLCQ $M$ theory compactified on a two-torus is
computed via supergravity.  We compare it to the effective action
obtained from the matrix theory, i.e., the (2+1)-dimensional
supersymmetric Yang-Mills (SYM) theory, including the one-loop
perturbative and full non-perturbative instanton effects.  Consistent
with the DLCQ prescription of $M$ theory {\em a la} Susskind, we find
the precise agreement for the finite $N$-sector (off-conformal
regime), as well as for the large $N$ limit (conformal regime),
providing us with a concrete example of the correspondence between the
matrix theory and the DLCQ $M$ theory.  Non-perturbative instanton
effects in the SYM theory conspire to yield the eleven-dimensionally
covariant effective action.
\vspace{2cm}
\end{titlepage}

%-------------------------------------------------------
\baselineskip 7mm

%\section{Introduction}

The original formulation of the matrix theory, a promising candidate
for the quantum description of $M$ theory, is given in the infinite
momentum frame (IMF) requiring us to consider the large parton number
limit \cite{bfss}.  Susskind then proposed the discrete light-cone
quantization (DLCQ) version of the matrix theory where one considers
the light-like eleventh direction, which was suggested to define the
matrix theory in the finite $N$-sector \cite{susskind}.  Becker,
Becker, Polchinski, and Tseytlin considered the scattering between two
$D$-particles ($M$-momentum) and showed that the matrix side
calculation for the effective action precisely reproduces the
eleven-dimensional supergravity side calculation up to two loops
\cite{becker}.  To find the precise agreement, it is indeed necessary
that the background geometry produced by the source $M$-momentum
should not be taken as the usual dimensionally uplifted version of the
$D$-particle solution along the spatial $M$ theory circle but as the
DLCQ version uplifted along the light-cone circle \cite{hks}. 
As advocated in \cite{hyun} and further elaborated in \cite{hyun2},
this tells us that the DLCQ $M$ theory should be compared with
the $M$ theory on a non-trivial background rather than
on a flat background.  Along this line of idea, the
same type of background geometry as in \cite{hks} was  
utilized in \cite{suss3} to show the
agreement between the DLCQ supergravity and the matrix theory for
three-graviton scatterings (see also \cite{etc}).

For the DLCQ $M$ theory compactified on a two-torus, the microscopic
description via the matrix theory is given by the $(2+1)$-dimensional
supersymmetric Yang-Mills theory \cite{watt,seiberg}.  Taking
the same DLCQ prescription for the background geometry of the
$N$-sector DLCQ $M$ theory on a two-torus yields a non-asymptotically
flat background space-time with the asymptotic geometry of the Anti de
Sitter (AdS) type \cite{hyun,hyun2}.  In the large $N$ and the large
eleventh radius $R$ (decompactification) limit while $N/R$ being
fixed, the background geometry becomes a tensor product of an AdS
space and a sphere, i.e., $AdS_4 \times S^7$.  In the same limit, the
matrix theory description turns to a conformal field theory (CFT) as
the infrared limit of SYM \cite{hyun2}.  This is the case where the
holographic AdS/CFT correspondence is conjectured to be valid
\cite{malda,witten}.  According to the motivation of the DLCQ $M$
theory \cite{susskind}, one might hope further that the correspondence
between the matrix theory and the DLCQ $M$ theory on the background
geometry of \cite{hyun,hyun2} persists even in the case of the finite
$N$ and the finite $R$.  In this paper, we find
an explicit example that directly supports this idea; we compute the
effective action for the membrane dynamics on the background geometry
of the DLCQ $M$ theory on a two-torus via supergravity for the finite
value $N$ and $R$.  We then show that the effective action up to the
fourth order in the membrane velocity exactly reproduces the effective
action computed from the (2+1)-dimensional SYM including the one-loop
perturbative and {\em full} non-perturbative instanton effects for the
finite value of $N$ and $R$, which were recently calculated in
\cite{sethi}.  Our effective action also behaves smoothly as we take
the limit where $N$ and $R$ goes to infinity while the ratio being
fixed.  Our analysis shows that the instantons in the matrix theory,
the (2+1)-dimensional SYM that does not have a manifest
eleven-dimensional covariance, conspire to yield the results that
derive from the manifestly covariant eleven-dimensional DLCQ
supergravity on the non-asymptotically flat background geometry.
    
In the case of the small value of $R$, the membrane dynamics in
string/supergravity theory and its comparison to the matrix theory
have been discussed in the literature \cite{malda2} - \cite{tsey2}.
In \cite{lif}, $D2$-brane scatterings in the
asymptotically flat ten-dimensional background geometry were found to
be in agreement with the perturbative one-loop SYM.  Polchinski and
Pouliot \cite{joe} added the one-instanton correction to the effective
action to investigate the effect of the $M$-momentum transfer and
found an approximate agreement with the eleven-dimensional
supergravity result.  The SYM side analysis was generalized in
\cite{dorey} to include the semi-classical multi-instanton effect.  A
clear physical interpretation of the instanton effects is also given
in \cite{kraus}.  In addition to the small $R$ limit, we note that by
adopting the asymptotically flat background geometry, it was necessary
to take the large $N$ limit to find the agreement in \cite{lif} -
\cite{kraus}, just as what happens in the IMF formulation of
\cite{bfss}.  In the context of the DLCQ $M$ theory, an agreement for
the finite $N$ and the small value of $R$ for the membrane dynamics
was observed in \cite{tsey2}.  In the limit of small $R$, the
instanton corrections are suppressed and the agreement in \cite{tsey2}
is up to the perturbative one-loop effect in SYM.  The salient feature
of our result presented here is the removal of the restriction to the
large $N$ and small $R$.  By adopting the background geometry of the
finite $N$-sector DLCQ $M$ theory proposed in \cite{hyun,hyun2} the
former restriction is removed.  Furthermore, by considering the full
non-perturbative instanton corrections on the SYM side, we are able to
find the agreement for an arbitrary value of $R$.  The conventional
range of the validity for the AdS/CFT correspondence is the large $R$
and the large $N$ limit while the ratio being fixed \cite{malda}.

%\section{DLCQ Supergravity}
   
The background geometry of $N$-sector DLCQ $M$ theory compactified on
a two-torus is given by the following eleven-dimensional metric
\cite{hyun,hyun2}
\begin{equation}
\label{m2metric}
ds^2_{11} = h^{-2/3} (-dt^2+dx_8^2+dx_9^2)
       +h^{1/3} (dx_1^2 + \cdots + dx_7^2+dx_{11}^2) , 
\end{equation}
where the eleven-dimensional harmonic function $h$ is given by 
\begin{equation}
h= \sum_{n = -\infty}^{\infty}
 \frac{\kappa N}{(r^2+( x_{11} + 2 \pi R  n)^2 ) ^3} . 
\label{m2harmonic}
\end{equation}
We introduce $r^2 = x_1^2 + \cdots + x_7^2 $ and the torus extends
over the $x_8$ and $x_9$ directions.  The integer $N$ represents the
number of the coincident source membranes and $\kappa$ is a dimensionful
constant.  The eleventh circle along the $x_{11}$ coordinate has the
radius $R$ and the function $h$ contains the contribution from all
possible mirror charges to respect the periodicity under the lattice
translation $x_{11} \rightarrow x_{11}+ 2 \pi R$.  In the limit of
large $R$, which corresponds to the decompactification limit of the
DLCQ $M$ theory, the metric (\ref{m2metric}) becomes that of the
$AdS_4 \times S^7$; the eleventh direction becomes indistinguishable
from other noncompact directions $( x_1 , \cdots , x_7 ) $ and the
transversal $SO(7)$ symmetry gets enhanced to the $SO(8)$ symmetry
\cite{nissan}.  Specifically, the summation in the expression for $h$
gets dominated by the $n=0$ term, which has the manifest $SO(8)$
symmetry.  This is the limit where we have the large $N$ AdS/CFT
correspondence, where the AdS supergravity and the CFT near the
infrared fixed point, i.e., the conformal phase of the
(2+1)-dimensional SYM, become a dual description to each other.  In
the limit of the vanishingly small $R$, we can replace the summation
in Eq. (\ref{m2harmonic}) with an integration and recover the
near-horizon geometry of the $N$ $D2$-branes of the type IIA
supergravity.  Our primary interest will be the case of the arbitrary
values of $R$ and $N$.  For the purpose of the comparison to the
(2+1)-dimensional SYM, which is the dimensional reduction of the
ten-dimensional SYM, it is convenient to reshuffle the series
summation of Eq.~(\ref{m2harmonic}) using the Poisson resummation
formula:
\begin{equation}
\sum^{\infty}_{n=-\infty} f(n) =
   \sum^{\infty}_{m=-\infty} \int^\infty_{-\infty} d \phi \;
   f( \phi ) \; e^{2 \pi i m \phi} ~.
\end{equation}
The resummation can be exactly performed to result the following
identity.
\[
  \sum_{n = -\infty}^{\infty}
  \frac{1}{(r^2+( x_{11} + 2 \pi R  n)^2 ) ^3}
 \  = \   \frac{1}{16 R} \ \Big[ \  \frac{3}{ r^5}       
  + \frac{1}{ r^3} 
      \sum^{\infty}_{m=1} \frac{m^2}{R^2} e^{-mr/R} 2 
        \cos (mx_{11}/R)  
\] 
\begin{equation}
      + \frac{3}{r^4} 
      \sum^{\infty}_{m=1} \frac{m}{R} e^{-mr/R} 2 
      \cos (mx_{11}/R) 
    + \frac{3}{r^5} 
      \sum^{\infty}_{m=1} e^{-mr/R} 2 \cos (mx_{11}/R)~ \Big] . 
     \label{resum} 
\end{equation}
We recognize the first term on the right hand side (RHS)  
as having the
structure of the perturbative one-loop term in the $D2$-brane
effective action \cite{lif}.  The summation index $m$ on the
RHS appears in the form of the $M$-momentum $m/R$.
According to the argument of Ref.~\cite{joe}, the integer $m$ may be
interpreted as the measure of the $M$-momentum transfer between the
source and probe membranes. The key observation for our later purpose
is that the three infinite series terms on the RHS can be rewritten in
terms of the modified Bessel function $K_\nu$ with a half-integer
$\nu$, which has the finite number of terms in an expansion
\cite{table}
\begin{equation}
\label{bessel}
K_{j+1/2} (z) = \left( \frac{\pi}{2 z} \right)^\frac{1}{2}
	e^{-z} \sum^j_{k=0} \frac{ (j+k)! }{ k! (j-k)! (2z)^k } ~.
\end{equation}
The function $h$ can then be succinctly written as
\begin{equation}
h(r, x_{11}) = \frac{\kappa N}{16 R} 
    \left[ \; \frac{3}{r^5} +
       \sum^{\infty}_{m=1} \left( \frac{2}{\pi} \right)^\frac{1}{2}
              \frac{m^2 m^{1/2}}{R^5} \left( 
         \frac{R}{r} \right)^{5/2}
                           K_{5/2} ( mr / R ) 
        2 \cos (m x_{11} / R ) \;
    \right] . 
\label{h}
\end{equation}

We now consider the dynamics of a probe membrane, which is taken to be
spanning the $x_8$, $x_9$ directions and is moving with a constant
velocity $v_i=\partial_0 x_i$ in a direction transversal to the probe
and $x_{11}$, i.e., $i=1,\cdots , 7$. The action for the probe
membrane is
\begin{equation}
\label{m2action}
S_2 = -T_2 \int d^3x \; \sqrt{ - \det h_{\mu \nu} }
                   + i \mu_2 \int H  ~,
\end{equation} 
where $T_2$ is the membrane tension and $H$ is the three-form gauge
field of the eleven-dimensional supergravity.  The metric $h_{\mu\nu}$
is the induced metric on the world-volume of the probe membrane given
by
\begin{equation}
h_{\mu\nu} = g_{\mu\nu} + \partial_\mu x^I \partial_\nu x^J g_{IJ}~.
\end{equation}
Here $\mu,~\nu$ are the world-volume indices and $I,~J$ are the
indices for the transverse directions to the probe.  We plug the
metric Eq.~(\ref{m2metric}) with the function $h$ of Eq.~(\ref{h})
into the action $S_2$ and expand it in powers of the transverse
velocity $v$. The action $S_2$ becomes
\begin{equation}
S_2 = \int d^3x \; \left[ \; \frac{1}{2} T_2 v^2
                 - V_2 + {\cal O} ( (v^2)^3 )
               \right] ~,
\end{equation}
where $V_2$ is the interaction potential given by
\begin{eqnarray}
V_2 &=& -\frac{1}{8} T_2 h(r,x_{11}) (v^2)^2 \nonumber \\
    &=& - \frac{N}{16 R M_p^{3}} (v^2)^2 \nonumber \\
    & & \times  \Bigg[ \;  \frac{3}{r^5} 
             + \sum^{\infty}_{m=1} 
               \left( \frac{2}{\pi} \right)^\frac{1}{2}
             \frac{m^2 m^{1/2}}{R^5} 
                   \left( \frac{R}{r} \right)^{5/2}
                     K_{5/2} (mr/R) 2 \cos (m x_{11}/R ) \;
        \Bigg] ~. \label{sugrapot}
\end{eqnarray}
Going to the last line, we use the fact that $T_2 \kappa = 8 M_p^{-3}$
where $M_p$ is the eleven-dimensional Planck scale \cite{joe}.  It
should be noted that the potential is valid for {\it any} value of
$R$.  If the value of $R$ is very small (or $r \gg R$) and we take
$m=1$, the potential is approximated by
\begin{equation}
V_2 \approx -\frac{3N}{16 R M_p^3} \frac{(v^2)^2}{r^5}
            -\frac{N}{16 R^3 M_p^3} \frac{(v^2)^2}{r^3} e^{-r/R} 
                 2 \cos (x_{11}/R)  ~.
\label{appr}
\end{equation}                                        
The first term on the RHS is the usual potential between two
$D2$-branes in the ten-dimensional type IIA theory \cite{lif} and
the second term is the potential due to the effect of a single
$M$-momentum transfer \cite{joe}. The approximate potential
Eq.~(\ref{appr}) shows a notable feature that there is no $r$
independent $v^4$ term that appeared in \cite{joe}.  Under the large
$N$ limit, it is natural to drop this term as was done in, for
example, \cite{lif}.  In the DLCQ framework, this term
is automatically absent \cite{hks,tsey2}, and this feature is
also present in the case of the exact potential, Eq.~(\ref{sugrapot}).

%\section{One-loop result in $(2+1)$-dimensional SYM}

According to the prescription of Seiberg and Sen, the DLCQ $M$-theory
on a two-torus is described by a system of $D2$-branes wrapped on its
$T$-dual two-torus \cite{seiberg}. When the number of
$D2$-branes is $N$, the action for the system is just the
(2+1)-dimensional $U(N)$ SYM theory. The interaction potential between
the source and the probe membranes is given by the effective potential
of the SYM theory, and we compare the supergravity effective potential
Eq.~(\ref{sugrapot}) to the effective potential of the SYM theory.  We
note that our supergravity side calculation is actually for the
two-body dynamics of the source and the probe.  This enables us to
restrict our attention only to the $SU(2)$ SYM theory, as far as the
dynamics is concerned.  From the gauge theory point of view, we do not
give the vacuum expectation values to the scalars that represent the
position of the $N$ source membranes, thereby making them localized at
one transversal space-time point.  In what follows, we thus put the
factor $N$ in front of the effective action and omit the trace part
\cite{lif,tsey2}.  We note that to tackle the genuine $N$-body
dynamics, we have to use the full $SU(N)$ SYM theory and this
problem is the beyond of scope of this paper.

Generically, the one-loop effective action $\Gamma^{(1)}$ of 
the SYM theory is
the summation of a perturbative term and $m$-instanton terms.
Hereafter, for the notational convenience, we call the perturbative
term $0$-instanton sector.  The general structure of the effective
action \cite{harvey,morales} looks schematically like
\begin{equation}
 \Gamma^{(1)} = N \int d^3 x \left( f^{(0)} u^4 
+ f^{(2)} u^3 \left[ \psi^2 \right] 
+ f^{(4)} u^2 \left[ \psi^4 \right]
+ f^{(6)} u \left[ \psi^6 \right]
+ f^{(8)}  \left[ \psi^8 \right] \right) , 
\end{equation}
where $u^i = \dot{\phi}^i= F_{0 i}$, the $i$-th component of the
electric field, $\left[ \psi^p \right] $ denotes a generic $p$ fermion
structure, and $f^{(p)}$ represents the bosonic coefficient function
of the corresponding $p$ fermion structure.  
The scalars $\phi^i$ ($i=1, \cdots , 7$) are the seven scalars of 
the vector multiplet (thereby having the $SO(7)$ symmetry). 
The function $f^{(p)}$
consists of the instanton summation and we represent it as $f^{(p)} =
\sum_{m=-\infty}^{\infty} f_m^{(p)} $ where $m$ is the instanton
number.  The bosonic zero fermion term $f^{(0)}$ should be compared to
what we found on the supergravity side analysis.  Recently, Paban,
Sethi and Stern \cite{sethi} exactly computed the eight fermion terms;
$[ \psi^8 ]$ consists of terms with zero, two and four scalar
structure, and they are determined via the
supersymmetry by requiring the absence of nine fermion terms under the
supersymmetry transformation.  Once $f^{(8)}$ term is calculated from
their analysis, the remaining terms are determined by the
supersymmetric completion.  To ensure the supersymmetry, we
have
\begin{equation} 
\delta \phi^i \left( \frac{\partial}{\partial \phi^i} 
f^{(p)} \right) \left[ \psi^p \right]_{\rm max} 
=  \delta \phi^i \phi^i \left(  \frac{d }
 {\phi d \phi}  f^{(p)} \right)  \left[ \psi^p 
   \right]_{\rm max} 
=  - f^{(p+2)} \delta_{\psi} \left[ \psi^{p+2} 
   \right]_{\rm max} ,
\end{equation}
where $\delta_{\psi}$ is only for the variation of fermion fields.  
Here, $\left[ \psi^p
\right]_{\rm max}$ denotes the $p$ fermion term with the maximum
number scalar structure.  We note that the function $f^{(p)}$ depends
only on the $SO(7)$ invariant combination $\phi^2 = \phi^i \phi^i$.
Thus, the function $f^{(0)}_m$ is related to $f^{(8)}_{4, m}$ by
\begin{equation}
\left( \frac{d}{\phi d \phi} \right)^4 f^{(0)}_m = k 
   f^{(8)}_{4,m} \  , 
\label{relat}
\end{equation}
where $f^{(8)}_{4, m}$ is the coefficient function of the four scalar
structure term among $\left[ \psi^8 \right]$ in 
the $m$-instanton sector
\begin{equation}
 \sum_{m= -\infty}^{\infty}
 f^{(8)}_{4,m} (\phi^i \phi^j \phi^k \phi^l T^{ijkl}) .
\end{equation}
Here $k$ is a constant and $T^{ijkl}$ is the eight fermion structure.
From \cite{sethi}, $f^{(8)}_{4,m}$ is given by
\begin{equation}
f^{(8)}_{4,m} =  m^6 |m|^{1/2} \frac{1}{g_{\rm YM}^{28}} 
	  \left( \frac{g_{\rm YM}^2}{\phi} \right)^{13/2}
	  K_{13/2} (|m| \phi / g_{\rm YM}^2 ) 
          e^{i m \phi^8 /g_{\rm YM}^2 } ~,
\label{8f}
\end{equation}
up to the overall multiplicative constant.  The extra scalar $\phi^8$
is the dual magnetic scalar and $g_{\rm YM}$ is the three-dimensional
Yang-Mills coupling constant.  We remark that Eq.~(\ref{8f}) gives the
perturbative term when we set $m=0$ proportional to $\phi^{-13}$.
Noting \cite{table} \begin{equation} \left( \frac{d}{z dz} \right)^a
(z^{-\nu} K_\nu (z)) = (-1)^a z^{-\nu -a} K_{\nu+a} (z)
\end{equation}
we conclude 
\begin{equation}
\label{nonpert}
f^{(0)}_m = C  m^2 |m|^{1/2} \frac{1}{g_{\rm YM}^{12}} 
	  \left( \frac{g_{\rm YM}^2}{\phi} \right)^{5/2}
	  K_{5/2} (|m| \phi / g_{\rm YM}^2) 
          e^{i m \phi^8 /g_{\rm YM}^2 }
\end{equation}
from Eq.~(\ref{relat}), where $C$ is a overall constant.  The constant
$C$ can not be determined by the argument of \cite{sethi}, but the
one-instanton calculation of \cite{joe} determines it to be $C=
(2/\pi)^{1/2} g_{\rm YM}^2/16$.  Thus, the bosonic one-loop effective
action $\Gamma^{(1)}_B$ from the SYM theory is
\begin{equation}
\Gamma^{(1)}_B =  N \int d^3x \sum_{m= -\infty}^{\infty}
 f^{(0)}_m (u^2)^2  \  , 
\label{symeff}
\end{equation}
including the full non-perturbative instanton corrections.  Since
$\Gamma^{(1)}_B = - \int d^3 x V_{\rm SYM}$, the effective potential
$V_{\rm SYM}$ is
\begin{eqnarray}
V_{\rm SYM} &=& - \frac{N}{16} (u^2)^2  \nonumber \\
        & \times & \left[ \; \frac{3}{\phi^5} +
               \sum^{\infty}_{m=1} 
                     \left( \frac{2}{\pi} \right)^\frac{1}{2}
                     \frac{m^2 m^{1/2}}{g_{\rm YM}^{10} }
             \left( \frac{g_{\rm YM}^2}{\phi}
                     \right)^{5/2} 
                     K_{5/2} ( m \phi / g_{\rm YM}^2 ) 
    2 \cos (m \phi^8 / g_{\rm YM}^2 ) \;
        \right] .
\label{sympot}
\end{eqnarray}
We note that Eq.~(\ref{sympot}) is exactly identical to
Eq.~(\ref{sugrapot}) if we identify $\phi_i = x_i / l_s^2 $, $\phi^8 =
x_{11} / l_s^2 $, $u = v / l_s^2$, and use $g_{\rm YM}^2 = g_s / l_s$.
The string coupling constant $g_s$ and the string length scale $l_s$
are related to the $M$ theory quantities by $g_s = (RM_p )^{3/2}$ and
$l_s = ( RM_p^3 )^{-1/2}$.

%\section{Discussions}
 
We thus constructed an explicit example that shows the agreement
between the matrix theory and the DLCQ $M$ theory on the 
non-asymptotically flat background
geometry of \cite{hyun}, which is consistent
with the finite $N$-sector DLCQ prescription of \cite{susskind}.  This
example points to the possibility that, at the level of finite $N$,
the AdS/CFT correspondence may be elevated to the dual description 
via the matrix theory of
the $M$ theory on the non-asymptotically-flat background geometry
of \cite{hyun}; a qualitative argument toward this effect has already
been given in \cite{hyun2}.

Regarding the matrix theory itself, our result clarifies
the space-time aspects; for example, the
locality along the $M$ theory circle and the eleven-dimensional
covariance are not at all manifest from the point of view of the
matrix theory.  Turning our calculations around, the effective action
of the (2+1)-dimensional SYM theory Eq.~(\ref{symeff}) can be
Poisson-resummed back to yield a eleven-dimensional harmonic
function proportional to Eq.~(\ref{m2harmonic}).  
At the level of the finite $R$, only
the $SO(7)$ symmetry is manifest.  As we take the limit where $R
\rightarrow \infty$ and $N \rightarrow \infty$ while keeping the 
ratio $N/R$ fixed (the IMF limit), however, 
Eq.~(\ref{m2harmonic}) shows that the symmetry
enhances to the $SO(8)$ symmetry ($n = 0$ term) and the
(2+1)-dimensional magnetic scalar combines with seven other scalars to
yield the eight coordinates transversal to the $M$ membranes.  This is
the expected covariance for the $M$ theory compactified on 
a two-torus.  To recover the eleven-dimensional covariance, the 
full non-perturbative
effects on the gauge theory side should be taken into account.


\begin{thebibliography}{99}
\bibitem{bfss} T. Banks, W. Fischler, S. H. Shenker, L. Susskind,
               Phys. Rev. {\bf D 55}, 5112 (1997).  
\bibitem{susskind} L. Susskind, hep-th/9704080.
\bibitem{becker} K. Becker, M. Becker, J. Polchinski, A. Tseytlin,
                Phys. Rev. {\bf D 56}, 3174 (1997), hep-th/9706072.
\bibitem{hks} S. Hyun, Y. Kiem and H. Shin, Phys. Rev. {\bf D 57},
              4856 (1998), hep-th/9712021.
\bibitem{hyun} S. Hyun, to appear in Phys. Lett. {\bf B}, 
               hep-th/9802026
\bibitem{hyun2} S. Hyun and Y. Kiem, hep-th/9805136.
\bibitem{suss3} J. McCarthy, L. Susskind and A. Wilkins, 
                hep-th/9806136.
\bibitem{etc} W. Taylor and M. Van Raamsdonk, hep-th/9806066;
	Y. Okawa and T. Yoneya, hep-th/9806108.
\bibitem{watt} W. Taylor, Phys. Lett. {\bf B 394}, 283 (1997), 
               hep-th/9611042.
\bibitem{seiberg} N. Seiberg, Phys. Rev. Lett. {\bf 79}, 3577 
                 (1997), hep-th/9710009;
              A. Sen, Adv. Theor. Math. Phys. {\bf 2}, 51 (1998),
              hep-th/9709220.
\bibitem{malda} J. Maldacena, hep-th/9711200. 
\bibitem{witten} S. Gubser, I. Klebanov and A. Polyakov, Phys. Lett. 
             {\bf B 428}, 105 (1998), hep-th/9802109;
             E. Witten, hep-th/9802150.
\bibitem{sethi} S. Paban, S. Sethi, and M. Stern, hep-th/9808119.
\bibitem{malda2} J. Maldacena, Nucl. Phys. Proc. Suppl.
       {\bf 68}, 17 (1998), hep-th/9709099. 
\bibitem{tsey} A. A. Tseytlin, Nucl. Phys. Proc. Suppl.
       {\bf 68}, 99 (1998), hep-th/9709123.
\bibitem{lif} G. Lifschytz and S. D. Mathur, Nucl. Phys. {\bf B499}, 
	283 (1997), hep-th/9612087;
        O. Aharony and M. Berkooz, Nucl. Phys. {\bf B491}, 184
        (1997), hep-th/9611215.
\bibitem{joe} J. Polchinski and P. Pouliot, Phys. Rev. D {\bf 56}, 
	6601 (1997), hep-th/9704029.
\bibitem{dorey} N. Dorey, V. V. Khoze, and M. P. Mattis, Nucl. Phys.
        {\bf B 502}, 94 (1997), hep-th/9704197.
\bibitem{kraus} E. Keski-Vakkuri and P. Kraus, hep-th/9804067.
\bibitem{tsey2} I. Chepelev and A. A. Tseytlin, Nucl. Phys.
        {\bf B 524}, 69 (1998), hep-th/9801120.
\bibitem{nissan} N. Itzhaki, J. Maldacena, J. Sonnenschein, and S. 
                 Yankielowicz, Phys.Rev. D {\bf 58}, 046004 (1998), 
                 hep-th/9802042.
\bibitem{table} I. S. Gradshteyn and I. M. Ryzhik, {\it Table of 
	Integrals, Series, and Products}, 5th ed. 
	(Academic Press, 1994).
\bibitem{harvey} J. A. Harvey, Nucl. Phys. Proc. Suppl. {\bf 68},
	113 (1998), hep-th/9706039.
\bibitem{morales} J. F. Morales, C. A. Scrucca, and M. Serone,
	hep-th/9801183.
\end{thebibliography}
\end{document}